\documentclass[12pt]{article}
\usepackage[dvips]{graphicx}
\usepackage{amssymb, amsmath, mathbbol}
\usepackage[centerlast,footnotesize]{caption2}

\newcommand{\gapprox}{\raisebox{-0.5ex}{$\
\stackrel{\textstyle>}{\textstyle\sim}\ $}}

\usepackage{graphicx}
\newcommand{\One}{1\kern-4.5pt1}

\newcommand{\be}{\begin{equation}}
\newcommand{\ee}{\end{equation}}

\def\lesim{${\lower 2pt\hbox{$\scriptstyle
<$}\atop\raise 4pt\hbox{$\scriptstyle\sim$}}$} 
\def\grsim{${\lower2pt\hbox{$\scriptstyle >$} \atop\raise4pt\hbox 
{$\scriptstyle\sim$}}$} 

\begin{document}
\begin{center}
\begin{flushright}
October 2009
\end{flushright}
\vskip 10mm
{\LARGE
Monte Carlo Simulation of the Semimetal-Insulator\\ 
\medskip
Phase Transition in Monolayer
Graphene
}
\vskip 0.3 cm
{\bf Wes Armour$^a$, Simon Hands$^b$ and Costas  Strouthos$^c$}
\vskip 0.3 cm
$^a${\em Diamond Light Source, Harwell Campus,\\
Didcot, Oxfordshire OX11 0DE, U.K.}
\vskip 0.3 cm
$^b${\em Department of Physics, Swansea University,\\
Singleton Park, Swansea SA2 8PP, U.K.}
\vskip 0.3 cm
$^c${\em Department of Mechanical Engineering, 
University of Cyprus,\\
Nicosia 1678, Cyprus.}
\vskip 0.3 cm
\end{center}

\noindent
{\bf Abstract:} 
 A 2+1 dimensional
fermion field theory is proposed as a model for the low-energy electronic
excitations in monolayer graphene. The model 
consists of $N_f=2$ four-component Dirac fermions moving in the plane and
interacting via a contact interaction between charge densities. 
For strong couplings there is a continuous transition to a Mott insulting phase.
We present results of an extensive numerical study of the model's critical region,
including the order parameter, its associated susceptibility,
and for the first time the quasiparticle propagator.  The data enables an
extraction of the critical exponents at the transition, including the dynamical
critical exponent, which are hypothesised
to be universal features of a quantum critical point. The relation of our model
with others in the literature is discussed, along with the implications
for physical graphene following from our value of the critical
coupling.

                                                                                

\section{Introduction}

There has been much recent interest in the
remarkable electronic properties of graphene
\cite{Novoselov:2005kj}
(see also a recent, comprehensive review 
in \cite{CNGPNG}). It appears that they arise from the
low-energy spectrum of excitations being equivalent to that of a two-dimensional 
gas of relativistic fermions, with the case of undoped (ie. neutral)
graphene corresponding to
zero net particle number. 
In brief, for a carbon monolayer with one mobile 
electron per atom,
a simple tight-binding model predicts 
a linear dispersion relation centred on zeroes located at the six 
corners of the first Brillouin
zone. 
It is possible to rewrite the Hamiltonian for single-particle excitations 
in Dirac form with $N_f=2$ flavors of 
four-component spinor $\psi$, the counting of 
degrees of freedom coming from 2
C atoms per unit cell $\times$ 2 zeroes per zone $\times$ 
2 physical spin components per electron. Electron propagation within the 
monolayer is thus relativistic, albeit with speed $v_F\approx c/300$. 

The charge carriers in graphene can be either electrons (``particles'') or
holes (``antiparticles'') and are characterised by a very high value of
mobility $\mu=\sigma/ne$ (where $\sigma$ is electrical conductivity and $n$ the
carrier density), more than twice that of the highest mobility conventional
semi-conductor, and several orders of magnitude greater than a typical metal at
room temperature.
This gives graphene the potential to be of great technological significance in
the construction of fast electronic devices. The naive tight-binding model
suggests, and 
experiments with graphene based on a SiO$_2$ substrate confirm, 
that graphene remains a conductor (technically a semimetal) for all values of
the gate voltage, ie. even when the carrier density formally vanishes, because
there is no gap in the energy spectrum at the Dirac points. The presence of a
small gap would, however, be extremely valuable for electronics applications, 
because it would increase the effective on-off current flow ratio needed for device
stability~\cite{CN}.

More sophisticated theoretical approaches to graphene must take inter-electron
interactions into account. In this paper we build upon an approach which
treats the low energy fermion excitations using 
a 2+1-dimensional relativistic quantum field
theory~\cite{Khveshchenko1,Gorbar:2002iw,Son:2007ja}. The interaction between
electrons is assumed
to be an unscreened Coulomb potential; because $v_F\ll c$ this can be treated as
``instantaneous'', meaning that the field theory is necessarily non-local.
The strength of the Coulomb interaction is variable, since it depends on the
dielectric constant $\varepsilon$
of the underlying substrate. It can be parametrised by an
effective fine structure constant
\begin{equation}
\alpha={e^2\over{4\pi\varepsilon\varepsilon_0\hbar v_F}}\sim O(1),
\end{equation}
so that the problem is strongly-interacting. The possibility then
opens up of the disruption of the free-field ground state by an excitonic
condensate, ie. one formed from tightly-bound electron-hole pairs, with
the effect of opening up a gap $\Delta>0$ at the Dirac point and making the ground
state of undoped graphene a Mott insulator.

The analogous phenomenon in particle physics is described using different language:
we say that the global chiral symmetry of the model, which prevents generation
of a fermion mass through quantum corrections to all orders in perturbation theory, is
spontaneously broken by the formation of a chiral condensate. Dynamical mass
generation is therefore inherently non-perturbative, and must be addressed
either by self-consistent analytic methods or, as in this paper, by numerical simulation
of a lattice-regularised version of the field theory. To date there have been
two distinct approaches taken. In a series of papers, Drut and
L\"ahde~\cite{D&L1,D&L2,D&L3} have simulated a formulation of the graphene field
theory based on lattice gauge theory, in which electrostatic degrees of
freedom are formulated on a $3+1$-dimensional lattice, while the electron fields
are restricted to a 2+1 dimensional slice. By contrast, our
formulation~\cite{Hands:2008id} is
entirely 2+1 dimensional, and is in essence a non-covariant form of the 
Thirring model. 
Both numerical calculations support the hypothesis proposed in
\cite{Son:2007ja} that the 
semimetal and insulator phases are separated by a line of second order 
phase transitions in the $(\alpha,N_f)$ plane, starting at a point
$(\infty,N_{fc})$ and running in the direction of decreasing $\alpha$ and
decreasing $N_f$. A very similar situation pertains in the 2+1$d$ Thirring
model~\cite{DelDebbio:1997dv,Christofi:2007ye}.
In \cite{Hands:2008id} we studied chiral symmetry breaking with variable $N_f$ in the
strong-coupling limit and
estimated $N_{fc}=4.8(2)$.
Each point on the line with integer $N_f<N_{fc}$ defines a quantum critical point (QCP)
whose properties are characterised by a set of $N_f$-dependent critical
exponents\footnote{in \cite{Son:2007ja} the identification of the QCP was restricted to
the case $(\infty,N_{fc})$.}.

In the current paper we present an extensive numerical study of the
semimetal-insulator phase transition for the case $N_f=2$ which is of direct physical
interest;
only preliminary results were available in
\cite{Hands:2008id}. The graphene model we study will be presented in detail, 
both as a continuum
field theory and as a lattice model, in the following
section, but it is appropriate to preface that with some remarks. Because our
approach is based upon a local quantum field theory in 2+1$d$, it is unable to
capture the long-range $1/r$ nature of the unscreened Coulomb potential assumed
in \cite{Khveshchenko1,Gorbar:2002iw,Son:2007ja}. We have argued in
\cite{Hands:2008id} that this is unimportant in the strong-coupling limit
$\alpha\to\infty$ where electron-hole pair polarisation effects dominate the
long-range physics; however for finite $\alpha$ our model is in principle
different both from the continuum approaches
\cite{Khveshchenko1,Gorbar:2002iw,Son:2007ja} and the lattice gauge theory
approach of \cite{D&L1,D&L2,D&L3}. We do not exclude the possibility, however,
that the universal behaviour at the QCP remains the same, even for $N_f<N_{fc}$.

There are two main benefits of our approach. Firstly, the simplicity of our
model and the fact that it is formulated directly on a 2+1$d$ spacetime lattice
mean that we have been able to perform accurate simulations on a range of system
sizes $L_s^2\times L_t$, yielding control over finite size artifacts and hence
access to the model's critical properties. Secondly,
the fact that our model is {\em not\/} a gauge theory permits a definition of the
quasiparticle correlation function without any need for gauge fixing, which is
known to be a major source of statistical noise in similar model systems,
eg.~\cite{Thomas:2006bj}. We are able here to present the first numerical study of
the quasiparticle propagator, which both explicitly demonstrates gap generation 
as the coupling strength $g^2$ is increased beyond $g_c^2$, 
and broadens the scope of the critical
analysis; for the first time we are able to present an estimate for the
dynamical critical exponent $z$ which governs the different scaling of the
correlation length in spacelike and timelike directions.

The remainder of the paper is organised as follows. In Sec.~\ref{sec:form} we
lay out the model to be studied in both continuum and lattice
formulations, and discuss its relation with other models studied in the
literature and its applicability to graphene. Our numerical results are
presented in Sec.~\ref{sec:results}: Sec.~\ref{sec:EoS} focusses on the chiral
order parameter and its associated susceptibility, 
and fits data to a renormalisation-group inspired
critical equation of state yielding estimates for the critical coupling and
exponents $\delta$ and $\beta$; Sec.~\ref{sec:quasi} presents an analysis of the
quasiparticle propagator and shows how both the gap $\Delta$ and the renormalised Fermi
velocity $v_{FR}$ may be extracted; finally Sec.~\ref{sec:z} presents a fit to a
similarly-motivated equation of state for $\Delta(m,g^2)$ which along with the
assumption of hyperscaling permits an estimate of the dynamical critcal exponent
$z$. We summarise our findings for the critical parameters in
Sec.~\ref{sec:discussion}, and also attempt to relate our value for
$g^2_c$ to estimates of $\alpha_c$ in the literature.

\section{Formulation and Interpretation of the Model}
\label{sec:form}

Our starting point is a model of relativistic Dirac fermions moving in 2+1
dimensions and interacting via an instantaneous Coulomb interaction. In
Euclidean metric the action is~\cite{Khveshchenko1,Gorbar:2002iw,Son:2007ja}:
\begin{equation}
S_1=\sum_{a=1}^{N_f}\int dx_0d^2x(\bar\psi_a\gamma_0\partial_0\psi_a
+v_F\bar\psi_a\vec\gamma.\vec\nabla\psi_a+iV\bar\psi_a\gamma_0\psi_a)
+{1\over{2e^2}}\int dx_0d^3x(\partial_i V)^2,
\label{eq:model}
\end{equation}
where $e$ is the electron charge, $v_F$ the Fermi velocity, $V$ the electrostatic potential, and the
$4\times4$ Dirac matrices satisfy $\{\gamma_\mu,\gamma_\nu\}=2\delta_{\mu\nu}$,
$\mu=0,1,2$. For monolayer graphene the correct number of fermion flavors
$N_f=2$. The momentum-space propagator for the $V$-field $D_1$, which couples
conserved charge densities $\bar\psi\gamma_0\psi$  at differing spacetime
points, is given by
\begin{equation}
D_1(p)
=\left({{2\vert\vec p\vert}\over e^2}+{N_f\over8}{{\vert\vec
p\vert^2}\over{(p^2)^{1\over2}}}\right)^{-1}
\label{eq:D_gr}
\end{equation}
where the first term in brackets on the right hand side is the classical Coulomb
interaction, and the second is the leading quantum correction in the large-$N_f$
limit, describing screening due to particle-hole virtual pairs. Note
that $p^2=p_0^2+v_F^2\vert\vec p\vert^2$. The relative importance of quantum
{\it versus\/}
classical effects may be parametrised by the ratio $\lambda$ of the two terms in
the static limit $p_0\to0$; in SI units
\begin{equation}
\lambda={{e^2N_f}\over{16\varepsilon\varepsilon_0\hbar
v_F}}\simeq{1.4N_f\over\varepsilon},
\label{eq:lambda}
\end{equation}
where $\varepsilon>1$ is the dielectric constant of the underlying substrate.

For sufficiently large interaction strength the description in terms of
massless relativistic excitations may be disrupted by condensation of bound
fermion-hole pairs in the ground state, signalled by an order parameter
$\langle\bar\psi\psi\rangle\not=0$, with the result that a gap appears in the
fermion spectrum. Physically this corresponds to a transition from a conductor
to an insulator; in the language of particle physics the same phenomenon,
resulting in a dynamical generation of a particle mass, is known as chiral
symmetry breaking. As this transition occurs at zero temperature, 
the model predicts a finite sequence of
quantum critical points (QCPs) whose properties at the critical interaction
strength $\lambda_c(N_f)$ are sensitive to
the value of $N_f$~\cite{Son:2007ja}: the sequence will terminate for
$N_{fc}$ (not necessarily integer) defined by $\lambda(N_{fc})=\infty$. 

This situation 
has motivated us to
explore a related but distinct model for graphene, with
action~\cite{Hands:2008id} (in units where $v_F=1$)
\begin{equation}
S_2
=\sum_{a=1}^{N_f}\int dx_0d^2x\left[\bar\psi_a\gamma_\mu\partial_\mu\psi_a
+iV\bar\psi_a\gamma_0\psi_a+{1\over{2g^2}}V^2\right]
\label{eq:ThirV}
\end{equation}
This model resembles the 2+1$d$ Thirring model~\cite{DelDebbio:1997dv,
Christofi:2007ye}, a four-fermi
model
known to exhibit a sequence of $N_f$-dependent QCPs as the coupling strength
$g^2$ is varied. The relation with (\ref{eq:model}) is clarified by inspection
of the $V$-propagator:
\begin{equation}
D_2(p)
=\left({1\over g^2}+{N_f\over8}{{\vert\vec
p\vert^2}\over{(p^2)^{1\over2}}}\right)^{-1}.
\label{eq:D_Th}
\end{equation}
Since the quantum correction is identical, models (\ref{eq:model}) and
(\ref{eq:ThirV}) should yield the same physics in the large-$\lambda$ limit,
which can be reached as either $N_f\to\infty$ or $g^2,e^2\to\infty$. Indeed, in
\cite{Hands:2008id} we used this property to predict the critical number of
flavors $N_{fc}=4.8(2)$ above which the model (\ref{eq:model}) remains a
semimetal for all $g^2$. Thus
graphene with $N_f=2$ is predicted to be a Mott insulator for sufficiently strong
inter-electron coupling.
For finite $N_f$ the models
(\ref{eq:model}) and (\ref{eq:ThirV}) are distinct, although we may still hope
they describe similar physics for $\lambda$ not too small. 

Let us discuss this point a little further. The principal difference 
between (\ref{eq:D_gr}) and (\ref{eq:D_Th}) occurs at large distances, ie. 
$\lim_{r\to\infty}D_1(r)\propto r^{-2}$, indicating that the long-range Coulomb
interaction is not screened, whereas $D_2$ is finite-ranged, being
cut off for $r\gapprox O(g^2)$. It is important to understand whether the
modification $D_1\mapsto D_2$ changes the physics in
any essential way, eg. by defining a model in a different universality class. We
will be unable to answer this question definitively with the simulation results
presented here, but note that in the model
approach of \cite{Gorbar:2002iw} which predicts dynamical symmetry breaking, the
relevant momentum range responsible for gap generation is $\vert\vec
p\vert\gg\Delta/v_F$, implying that it is the short-ranged behaviour of $D$
which governs the properties of the QCP. In addition, we note that unlike the
``instantaneous'' approximation used in that work $D_2$ correctly incorporates
the $p_0$-behaviour of the vacuum polarization function.

In this paper we will use numerical simulations of a lattice model based opon a
discretised version of (\ref{eq:ThirV}) to study the semimetal-insulator
transition for the physical value $N_f=2$.
The lattice model 
is formulated in terms
of 
single-component Grassmann fields $\chi,\bar\chi$ defined on the sites
$x$ of a three-dimensional cubic lattice, by the action
\begin{equation}
S_{latt}=\sum_{x\mu}\bar\chi_{x}{\eta_{\mu x}\over2}
\bigl[(1+\delta_{\mu 0}\sqrt{2g^2}e^{iV_x})
\chi_{x+\hat\mu}-
(1+\delta_{\mu 0}\sqrt{2g^2}e^{-iV_{x-\hat0}})
\chi_{x-\hat\mu}\bigr]
+m\sum_{x}\bar\chi_{x}\chi_{x}.
\label{eq:latt}
\end{equation}
The sign factors
$\eta_{x\mu}\equiv(-1)^{x_0+\cdots+x_{\mu-1}}$ ensure that in the
long-wavelength limit the first (antihermitian) term in $S_{latt}$ describes
the Euclidean propagation of $N_f=2$ flavors of relativistic fermion described
by four-component spinors
\cite{BB}. The bare fermion mass 
$m$ 
provides a IR regulator for modes which would otherwise be massless
in the limit of weak interactions. 
The hopping terms in $S_{latt}$ 
involve the auxiliary boson field $V_x$ which is formally defined on the {\it
timelike\/} links connecting sites $x$ with $x+\hat0$. 
For further details of the relation  between the actions (\ref{eq:latt}) and
(\ref{eq:ThirV})
we refer the reader to \cite{DelDebbio:1997dv} and \cite{Hands:2008id}
\footnote{In particular, for $N_f=2$ the action (\ref{eq:latt})
can be recast in a ``non-compact'' form yielding identical physics,
whose relation to to (\ref{eq:ThirV}) is
more manifest.}. 
Because we restrict our consideration to an integer number of fermion flavors
in this paper we were able to perform the 
simulation using a well-established numercial method
called the hybrid Monte Carlo (HMC) algorithm, which generates
equilibrated ensembles of field configurations $\{V\}$ with no systematic
bias~\cite{HMC,DelDebbio:1997dv}. 

Since our initial paper \cite{Hands:2008id}, results from
simulation of an alternative
lattice approach to the graphene model (\ref{eq:model}) have 
appeared~\cite{D&L1,D&L2,D&L3}.
This formulation is based on lattice gauge theory, in which the
degrees of freedom  
corresponding to the electrostatic potential $V$ are formulated on a
3+1$d$ lattice, while the electron degrees of freedom are confined to a 2+1$d$
``braneworld''. This in principle gives a more faithful rendering of the 
physics encapsulated in (\ref{eq:model},\ref{eq:D_gr}). 
The principal result is a prediction for the
critical coupling corresponding to the semimetal-insulator transition; for
$N_f=2$ the value
\begin{equation}
\lambda_c^{DL}=1.70(2)
\end{equation}
was obtained.
This result is intriguing because it lies between the values $\lambda\simeq1.25$
expected for graphene on an SiO$_2$ substrate, which experiments have shown
to be a conductor, and $\lambda\simeq3.4$ (Cf. eqn.~(\ref{eq:lambda})) 
for freely-suspended graphene,
which is accordingly predicted to be an insulator. The necessity to store and
evolve variables on an extra dimension is clearly a computational burden the
action (\ref{eq:latt})
evades; however for current purposes the main advantage we claim for
our approach is that it permits a straightforward means to measure the
quasiparticle propagator, as presented below in Sec.~\ref{sec:quasi}, 
without the need for gauge
fixing.

Let us finish this section by discussing the behaviour we expect of the model
(\ref{eq:latt}), and how it might relate to physical graphene.  In the limit
$m\to0$ the model has a global chiral symmetry $\chi_x\mapsto\exp
i\alpha\varepsilon_x\chi_x; \bar\chi_x\mapsto\exp i\alpha\varepsilon_x\bar\chi_x$
where the sign factor $\varepsilon_x\equiv(-1)^{x_0+x_1+x_2}$ distinguishes 
odd and even sublattices: 
the model studied in ~\cite{D&L1,D&L2,D&L3} has the identical symmetry.
For $N_f$ flavors 
the pattern of symmetry breaking expected for the continuum models
(\ref{eq:model}, \ref{eq:ThirV}) is U($2N_f)\!\!\to$U$(N_f)\otimes$U$(N_f)$, whereas
away from the continuum limit the pattern for (\ref{eq:latt}) is
U$({N_f\over2})\otimes$U$({N_f\over2})\!\!\!\to$U$({N_f\over2})$.
By analogy with
the Thirring model \cite{DelDebbio:1997dv}, we expect that for large values of
the coupling $g^2$ the symmetry will be dynamically broken as signalled by a 
non-vanishing condensate $\langle\bar\chi\chi\rangle\equiv
V^{-1}\partial\ln{\cal Z}/\partial m\not=0$, but that the
symmetry will be restored in a continuous phase transition at some critical
coupling
$g_c^{-2}$. This transition between the two phases defines a UV-stable fixed point of the
renormalisation group, and the fixed point theory is thus uniquely specified by a
set of critical exponents -- one of the main goals of the paper, presented in
Sec.~\ref{sec:EoS}, is to
determine both the critical $g_c^{-2}$ and the set of exponents by numerical
means. The fixed-point theory should describe the low-energy excitations of
physical graphene in the continuum limit, reached either from the
insulating phase as $g^{-2}\nearrow g_c^{-2}$, $\langle\bar\chi\chi\rangle\to0$,
or from the conducting phase as $g^{-2}\searrow g_c^{-2}$, $m\to0$. 
As we shall see below, it may be possible to relate the value of  $g_c^{-2}$ to 
$\lambda$ as defined via (\ref{eq:lambda}). The
applicability of the model to graphene, however, rests on the hypothesis that 
the low-energy excitations and their interactions in graphene share its symmetries, and that 
the physical parameters are such that graphene lies within the basin of
attraction of the fixed point. Ultimately this must be settled by experiment.

\section{Numerical Results}
\label{sec:results}

Preliminary simulations with $N_f=2$ presented in \cite{Hands:2008id} showed
evidence for a crossover from strong- to weak-coupling behaviour at
$g^{-2}\approx0.6$. Accordingly, we have undertaken a refined campaign of
simulation on system sizes $L_s^2\times L_t$ with $L_s$ ranging from 16 to 48,
and $L_t$ ranging from 48 to 84; the bare mass $m$ was varied between 0.0025 and
0.025. 
Because the action (\ref{eq:latt}) does not treat space and (Euclidean) time directions
equivalently, it is useful to explore the consequences of
independently varying $L_s$ and $L_t$: however,  
the most detailed coverage of the $g^{-2}$ axis was obtained on
$24^2\times48$.

\subsection{Equation of State}
\label{sec:EoS}

In the vicinity of a second order phase transition the order parameter 
over a range of couplings $g^{-2}$ and small 
source values $m$ can be described by an
equation of state of the form
\begin{eqnarray}
m&=&\langle\bar\chi\chi\rangle^\delta{\cal
F}(t\langle\bar\chi\chi\rangle^{-{1\over\beta}})\nonumber\\
&=&A
t\langle\bar\chi\chi\rangle^p
+B\langle\bar\chi\chi\rangle^\delta+O
\bigl(t^2\langle\bar\chi\chi\rangle^{p-{1\over\beta}}\bigr),
\label{eq:eos}
\end{eqnarray}
where 
$g_c^{-2}$ is the critical coupling, $t\equiv g^{-2}-g_c^{-2}$,
${\cal F}$ is a universal scaling function, $\delta$ the critical exponent
describing the order parameter's response at criticality to a small applied
source $m$, $\beta$ the exponent governing the
scaling of the order parameter for $m=0$ as $t\to0_-$,
and $p\equiv\delta-1/\beta$.  
Order parameter data taken in the thermodynamic limit can be
fitted to (\ref{eq:eos}) to extract $g_c^{-2}$, $\delta$ and $p$. In practice we
need to make assumptions about the width of the ``scaling window'' in $g^{-2}$
and $m$ where the subleading
corrections in (\ref{eq:eos}) can be safely ignored, 
and we also need to carefully monitor the effects of working with
finite $L_s$, $L_t$.

\begin{figure}[tb]
\begin{center}
\begin{minipage}[c][7.8cm][c]{7.4cm}
\begin{center}
    \includegraphics[width=7.4cm]{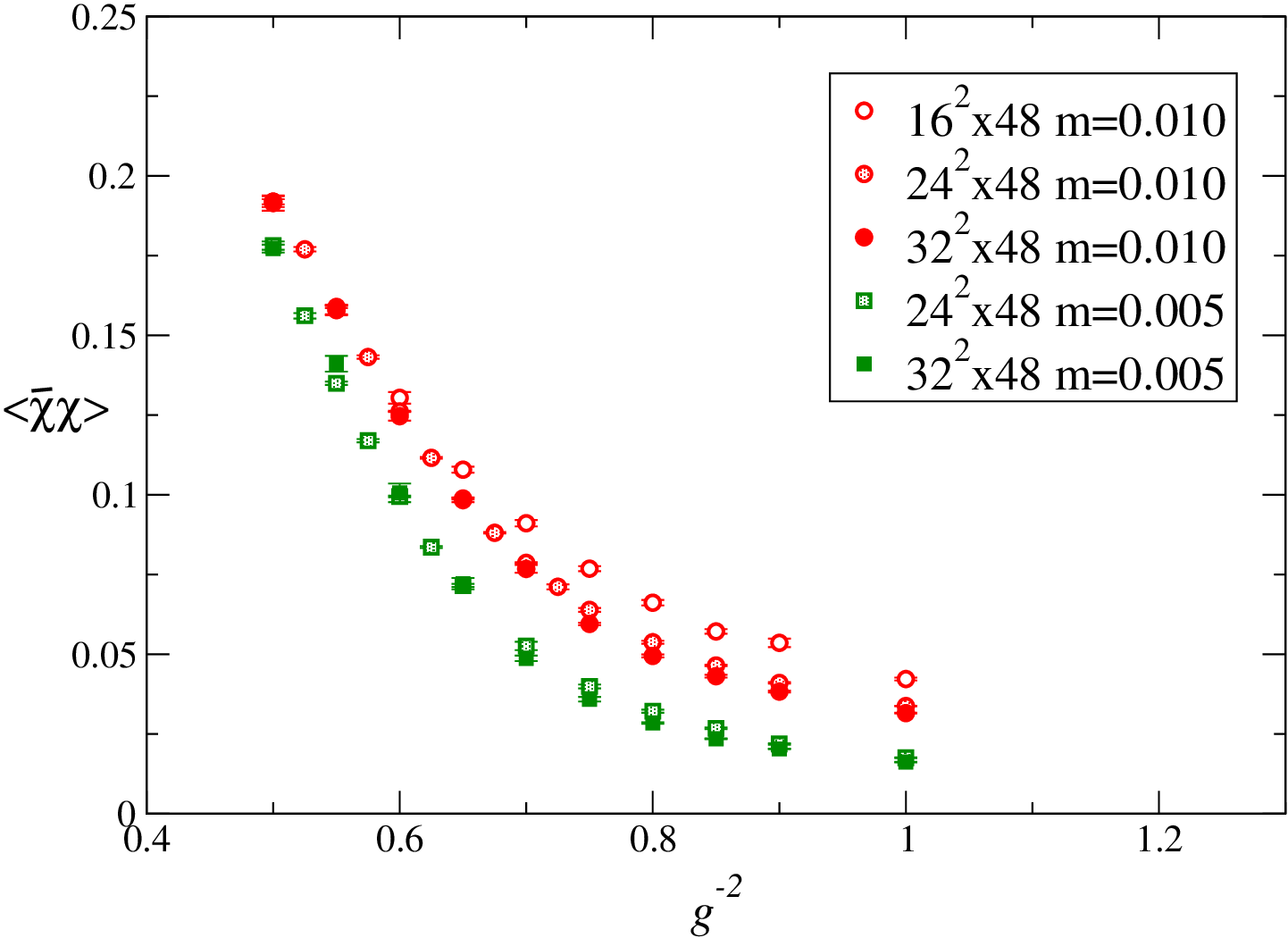}
\\[-1mm]
(a)
\end{center}
\end{minipage}
\begin{minipage}[c][7.8cm][c]{7.4cm}
\begin{center}
    \includegraphics[width=7.4cm]{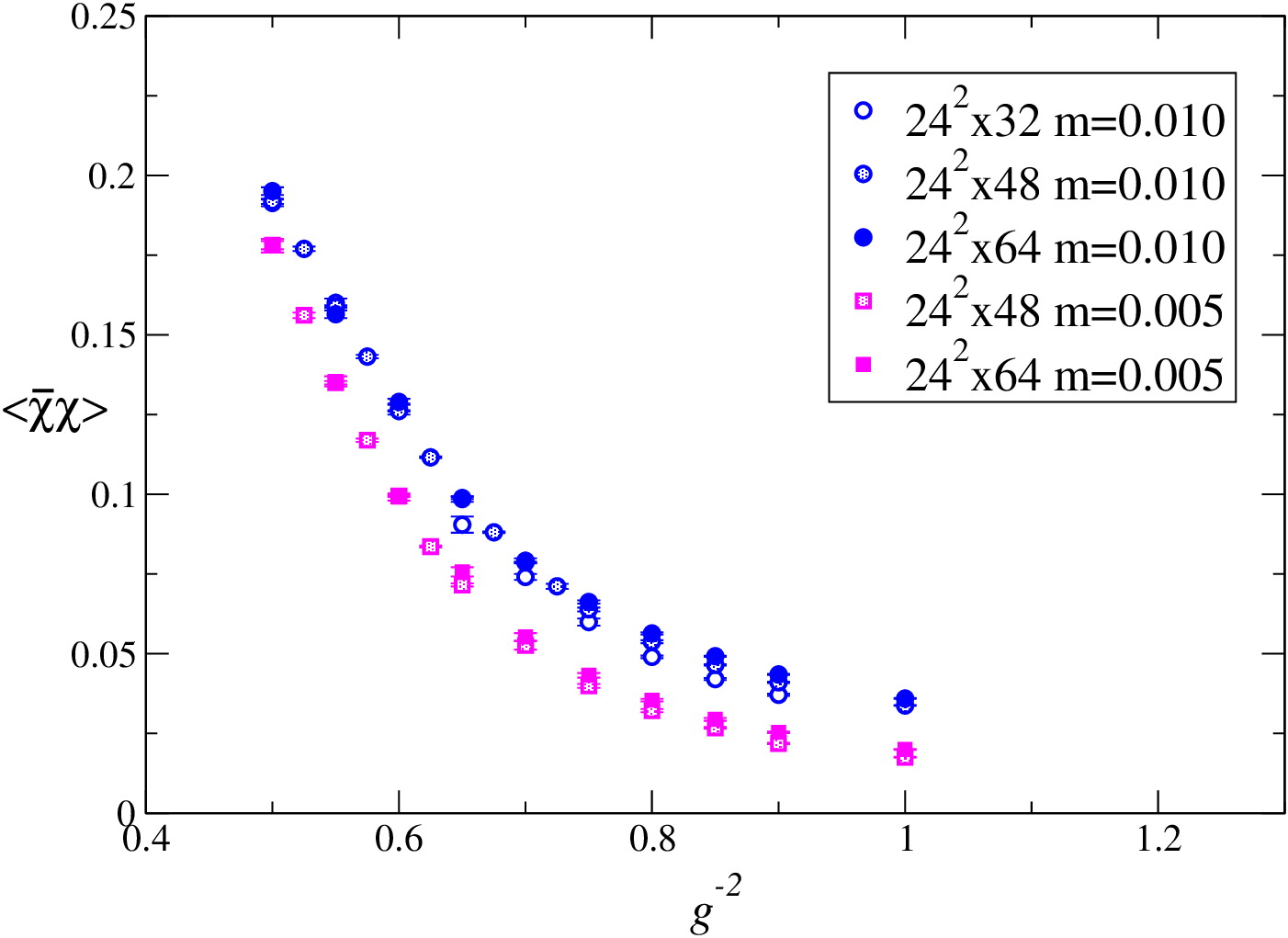}
\\[-1mm]
(b)
\end{center}
\end{minipage}
\caption{(Color online) 
$\langle\bar\chi\chi\rangle$ vs. $g^{-2}$ for $m=0.01$ calulated on
(a) $L_s^2\times48$ lattices;
(b) $24^2\times L_t$ lattices, showing finite size effects.} 
\label{fig:finite_size}
\end{center}
\end{figure}

First let's discuss finite volume effects. Since the 
model (\ref{eq:latt}) has an anisotropic action,
we cannot {\it a priori\/} exclude the possibilty of
correlation lengths in spatial and temporal directions diverging with distinct
exponents $\nu_s$ and $\nu_t$~\cite{BW}. In previous work we have
attempted to incoporate this possibility via a correction to the equation of
state fit, but the complicated nature of the finite volume scaling model
made these fits of questionable value given the range of simulation volumes
available to us. Here we take a more pragmatic approach, and compare order
parameter data for two different bare masses at
fixed $L_t$ and varying $L_s$ in Fig.~\ref{fig:finite_size}a and {\it vice
versa\/} in Fig.~\ref{fig:finite_size}b. The plots reveal the very different
nature of the finite size effects in each case: $\langle\bar\chi\chi\rangle$
rises as $L_t$ is increased, corresponding to the zero temperature limit, 
but falls as the thermodynamic limit $L_s\to\infty$ is approached. 
Moreover, in both cases the
effects are greater in the symmetric phase, ie. at larger values of $g^{-2}$.
We will proceed by using the observation that in the restricted range $0.525\leq
g^{-2}\leq0.65$ the data for $m=0.005$ 
($0.525\leq g^{-2}\leq0.70$ for $m\geq0.01$) on $24^2\times48$ 
are free from finite size effects almost within
statistical error.

\begin{figure}[htbp]
    \centering
    \includegraphics[width=13.0cm]{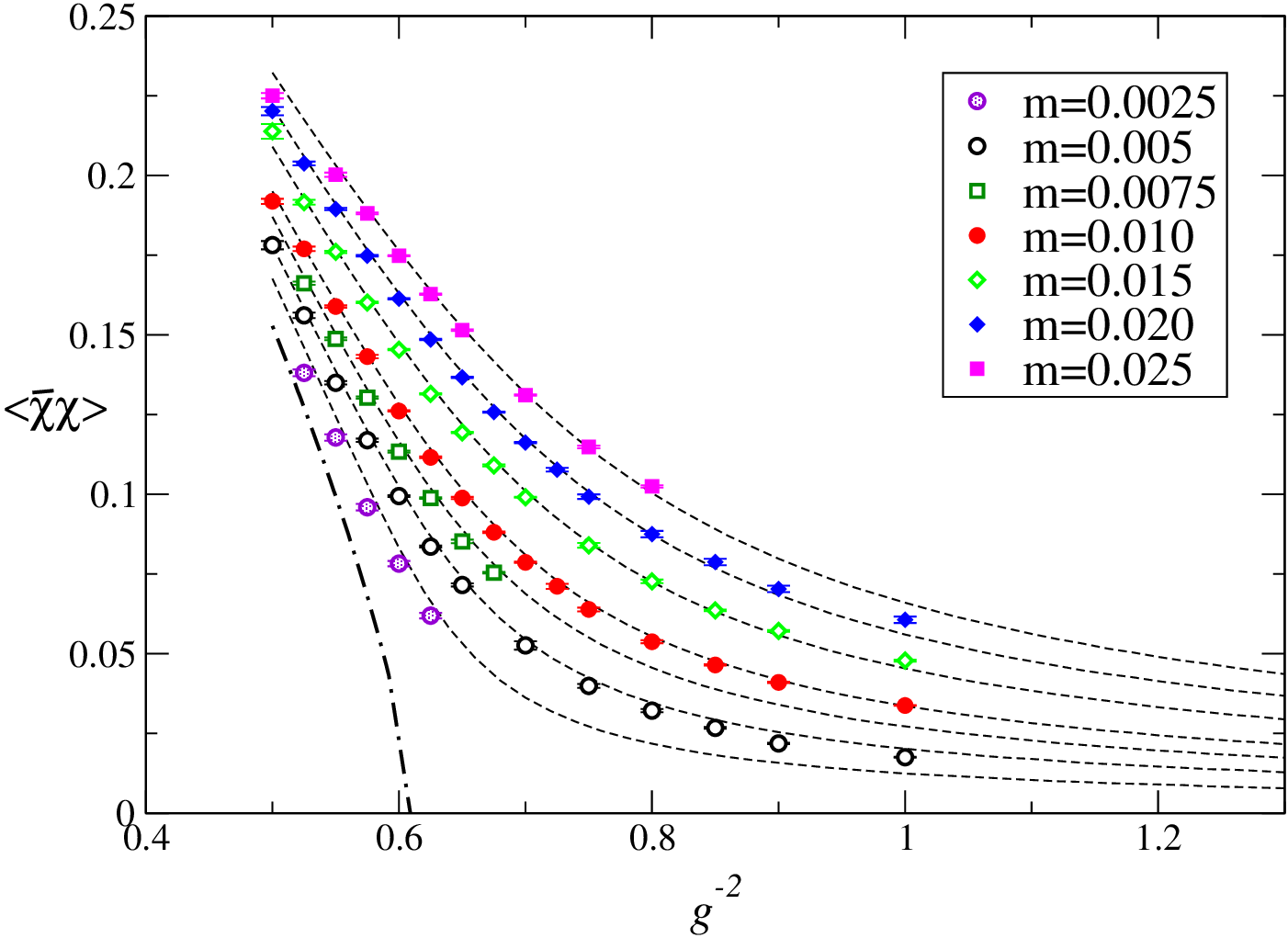}
    \caption{(Color online) 
Fit to (\ref{eq:eos}) to order parameter data taken on $24^2\times48$.
The function in the $m\to0$ limit is also shown.} 
   \label{gr:power}
\vspace{8mm}
    \centering
    \includegraphics[width=13.0cm]{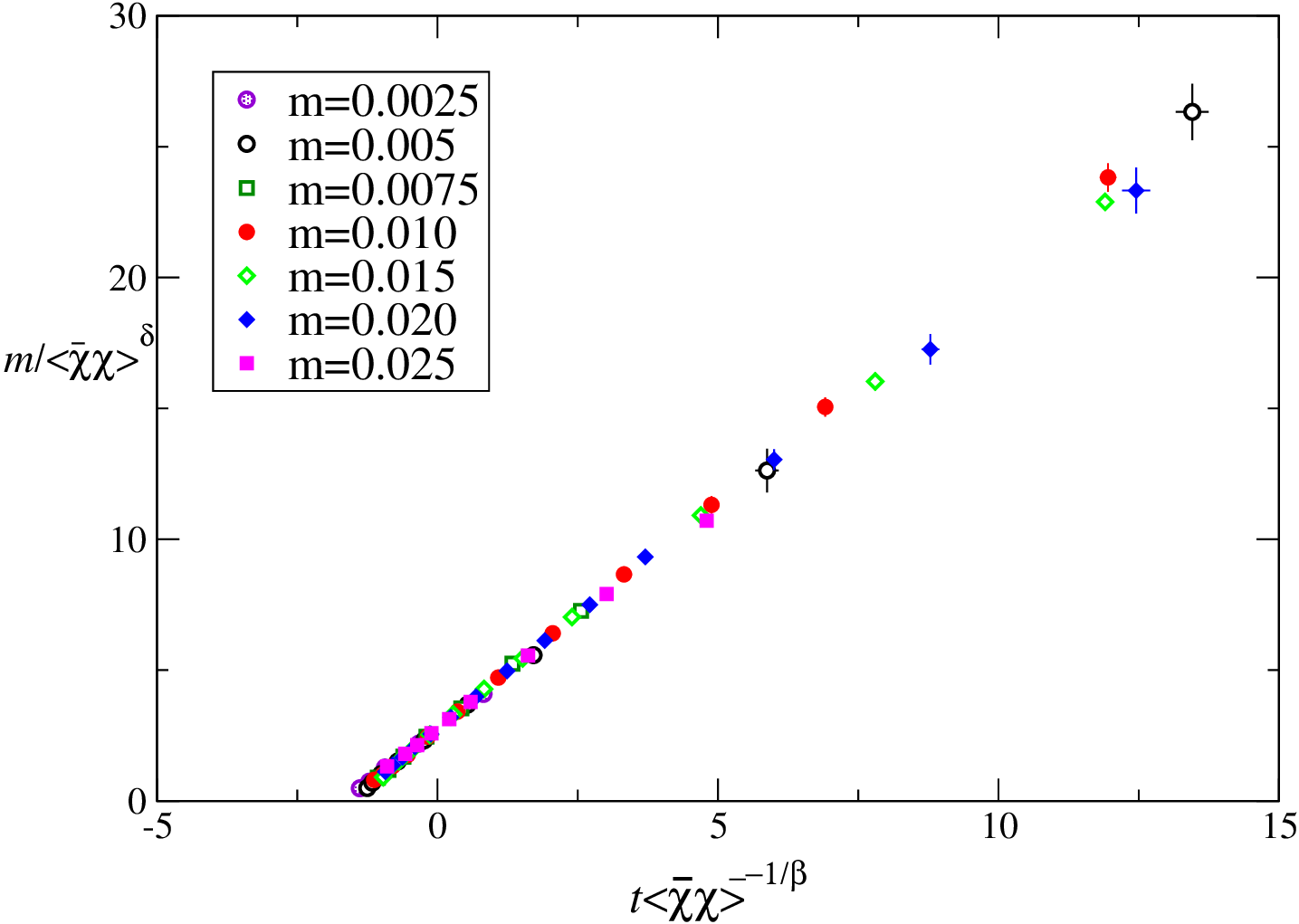}
    \caption{(Color online) 
Plot of $m/\langle\bar\chi\chi\rangle^\delta$ vs.
$(g^{-2}-g_c^{-2})\langle\bar\chi\chi\rangle^{-1/\beta}$ using the critical
parameters (\ref{eq:fits}).}
   \label{gr:scaling}
\end{figure}

\begin{table}[h]
\centering
\setlength{\tabcolsep}{0.4pc}
\hspace{-20mm}
\begin{tabular}{|ll|lll|l|}
\hline
fit & \# & $g_c^{-2}$ & $\delta$ & $p$ & 
$\chi^2$/dof \\
\hline
$0.525\leq g^{-2}\leq0.90$ (all $m$) & 69 & 0.608(2) & 2.66(2) 
& 1.252(4) &  6.9\\
&&&&&\\
\hline
$0.55\leq g^{-2}\leq0.80$ (all $m$)& 55 & 0.607(2) & 2.68(3) 
& 1.261(9) &  4.0\\
&  &  &  &  &  \\
\hline
$0.525\leq g^{-2}\leq0.65$ ($m=0.005$)& 43 & 0.609(2) &
2.66(3) 
& 1.245(11) &  2.7\\
$0.525\leq g^{-2}\leq0.675$ ($m=0.0075$)&  &  &  &  &  \\
$0.525\leq g^{-2}\leq0.70$ ($m\geq0.01$)&  &  &  &  &  \\
\hline
$0.525\leq g^{-2}\leq0.70$ ($m\geq0.0075$)&  
 37 & 0.600(3) & 2.80(5) 
& 1.285(14) &  2.3\\
\hline
\end{tabular}
\caption{Various fits to the Equation of State (\ref{eq:eos}) for data taken on
$24^2\times48$} 
\smallskip
\label{tab:eos}
\end{table}
Table~\ref{tab:eos} shows sample fits to (\ref{eq:eos}) to $O(t)$
for order parameter data
taken on a $24^2\times48$ lattice, and shows how the fit quality improves as
data far from criticality are successively excluded. We also tried excluding low
mass points as these are most susceptible to finite volume effects. 
It is comforting
that the fitted values of the critical parameters are quite stable as the
scaling window is so varied. Our
preferred fit is the third row of Table~\ref{tab:eos}, 
which includes as much data as possible consistent with
preserving acceptable fit quality, is
\begin{equation}
g_c^{-2}=0.609(2);\;\;\;\delta=2.66(3);\;\;\;p=1.245(11)\;\;\;
\Rightarrow\;\;\;\beta=0.71(2).
\label{eq:fits}
\end{equation}
The fitted equation of state is plotted in Fig.~\ref{gr:power}.

We also experimented with fits with an extra free parameter modelling an
$O(t^2)$ correction to (\ref{eq:eos}); these fits indicated a slightly larger
value of $g_c^{-2}$, but despite the extra free parameter did not yield
appreciably better $\chi^2$ values. Moreover, the resulting equation of state
clearly failed to be physically reasonable in the symmetric phase: since 
$\delta-{2\over\beta}<0$ from (\ref{eq:fits}), the $O(t^2)$ term rapidly becomes
numerically dominant here.
In support of this, Fig.~\ref{gr:scaling} plots order
parameter data taken on $24^2\times48$ using
axes chosen, using the
critical parameters (\ref{eq:fits}), to effect data collapse onto the scaling
function ${\cal F}$ which is seen to be linear to very good approximation.

Finally we examine another probe of the critical point, the ratio
of transverse to longitudinal susceptibilities
$\chi_\ell/\chi_t\equiv\partial\ln\langle\bar\chi\chi\rangle/\partial\ln
m$~\cite{D&L2,DelDebbio:1997dv}.
There are two spin-0 particle-hole channels with opposite intrinsic parities,
which by analogy with mesons in particle physics we refer to as $\sigma$ (parity +) and
$\pi$ (parity $-$). In the $m\to0$ limit, in the conducting phase the two states
are related by a U(1) global chiral symmetry and are degenerate; in the insulating
phase, by contrast, the symmetry is spontaneously broken and the $\pi$-channel
therefore contains a massless pole by Goldstone's theorem. Since
$\chi_\ell/\chi_t$ is simply the ratio of the integrated $\sigma$-propagator to
the integrated $\pi$-propagator, we expect it to tend to unity as $m\to0$
in the conducting phase, and to zero in the insulating phase. Exactly at
criticality, however, the ratio is $m$-independent and takes the value
$1/\delta$~\cite{Kocic:1992pf}.
\begin{figure}[ht]
    \centering
    \includegraphics[width=13.0cm]{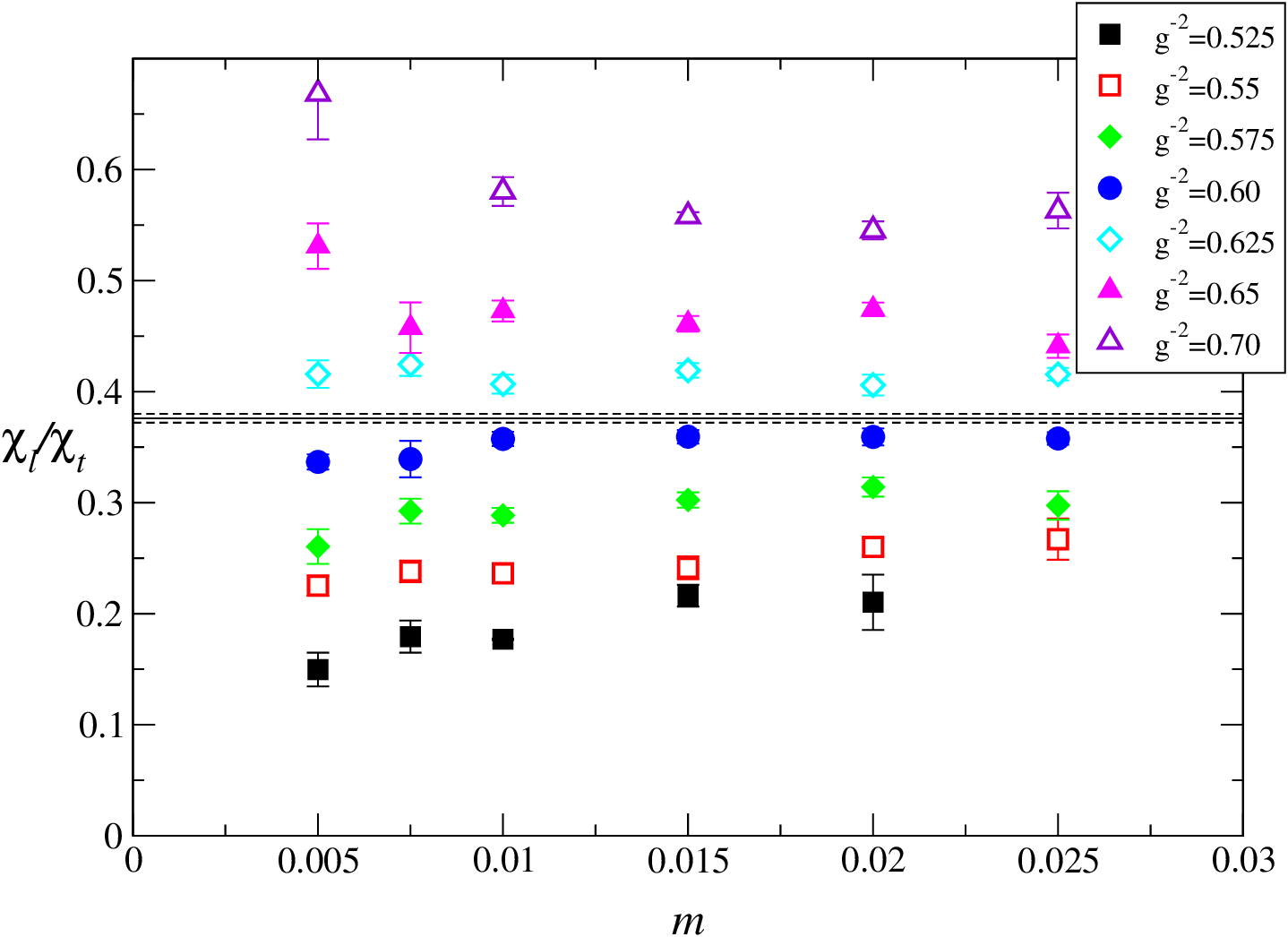}
    \caption{(Color online) 
Susceptibility ratio $\chi_\ell/\chi_t$ vs. $m$ for various $g^{-2}$ in the
critical region. The fitted value of $\delta^{-1}$ from (\ref{eq:fits}) is shown
as a horizontal band.}
   \label{gr:suscrat}
\end{figure}
Fig.~\ref{gr:suscrat} plots $\chi_\ell/\chi_t$ vs. $m$ evaluated on a
$24^2\times48$ lattice, including contributions from diagrams with both
connected and disconnected fermion lines as detailed in \cite{DelDebbio:1997dv}.
The data
taken at $g^{-2}=0.60, 0.625$ are approximately $m$-independent, especially for
larger $m$, and
bracket the value of 
$\delta^{-1}$ obtained from the equation of state fit, 
strengthening our confidence in the values of the critical parameters in
(\ref{eq:fits}).

\subsection{Quasiparticle Dispersion}
\label{sec:quasi}

One of the main motivations for the choice of model (\ref{eq:latt}) is that
since it has no manifest gauge symmetry, there is no requirement to fix a gauge
in order to define or measure a correlation function such as the fermion
propagator. This has enabled us to perform the first 
numerical simulation of the quasiparticle excitation spectrum in graphene.

The fermion excitation spectrum of the model is accessed via analysis of the
Euclidean timeslice propagator $C_f(\vec p,t)$ defined by
\begin{equation}
C_f(\vec p,t)=
\sum_{\vec x\;{\rm even}}\langle\chi(\vec0,0)\bar\chi(\vec x,t)\rangle
e^{-i\vec p.\vec x},
\end{equation}
where ``even'' refers to sites with spatial coordinate $\vec x$ obeying
$(-1)^{x_1}=(-1)^{x_2}=1$, and the components of $\vec p$ take values $2\pi
n/L_s$, with $n=0,1,\ldots,L_s/4$. 
This restriction improves the signal to noise ratio, and
originates in the observation that the action (\ref{eq:latt}) is invariant only
under translations by an even number of lattice spacings.
The energy $E(\vec p)$ is then
extracted by a fit of the form
\begin{equation}
C_f(\vec p,t)=B(e^{-Et}+e^{-E(L_t-t)}),
\label{eq:Efit}
\end{equation}
where in this case only data with $t$ odd were used, since this yielded the
best fits across the whole range of $g^{-2}$ (it can be shown that
$\lim_{m\to0}C_f=0$
for even $t$ in the conducting phase).

We measured $E(\vec p)$ for $\vec p=(p_1,0)$ on $32^2\times48$ for
$g^{-2}=0.55,0.6,0.7,0.8$, and additionally on $48^3$ for $g^{-2}=0.8,0.9$,
using $m$ ranging from 0.005 to 0.03.
\begin{figure}[ht]
    \centering
    \includegraphics[width=13.0cm]{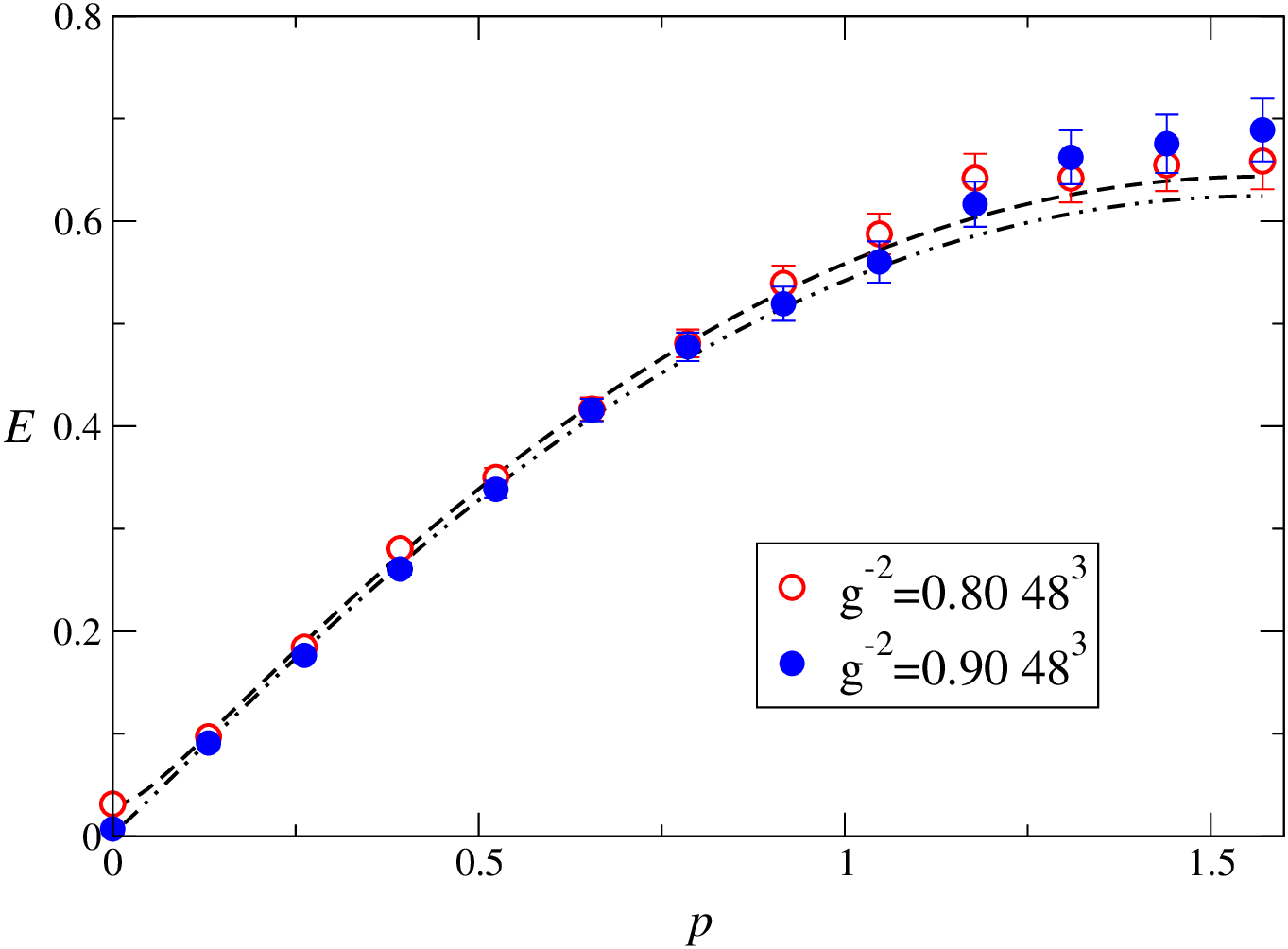}
    \caption{(Color online) 
Dispersion relation $E(p)$ as measured on a $48^3$ lattice with $m=0.005$.}
   \label{gr:dispersion}
\end{figure}
The resulting dispersions for the latter two systems at $m=0.005$ are shown in
Fig.~\ref{gr:dispersion}. For small $p$ and $m$ the dispersion starts out linear to good
approximation, and then flattens out to have zero slope at the 
effective Brillouin zone edge at $p={\pi\over2}$; this flattening is a
discretisation artifact with no physical significance.
To proceed we parametrise the dispersion relation
using 
\begin{equation}
E(p)=A\sinh^{-1}(\sqrt{\sin^2p+M^2}),
\label{eq:dispfit}
\end{equation}
where for $A=1$ and $M=m$ the exact result for non-interacting
lattice fermions is recovered.
Two sample fits are shown in Fig.~\ref{gr:dispersion}.
For small $M$ we can interpret $E(0)\equiv\Delta\approx AM$ as the quasiparticle mass
(or gap), and for small $p$ 
in the limit $M\to0$ then $dE/dp\approx A$ is the physical Fermi velocity
$v_{FR}$. 
Results for $A$ and $M$ as functions of $m$ are shown in
Figs.~\ref{gr:M},\ref{gr:A}.
\begin{figure*}[htbp]
    \centering
    \includegraphics[width=13.0cm]{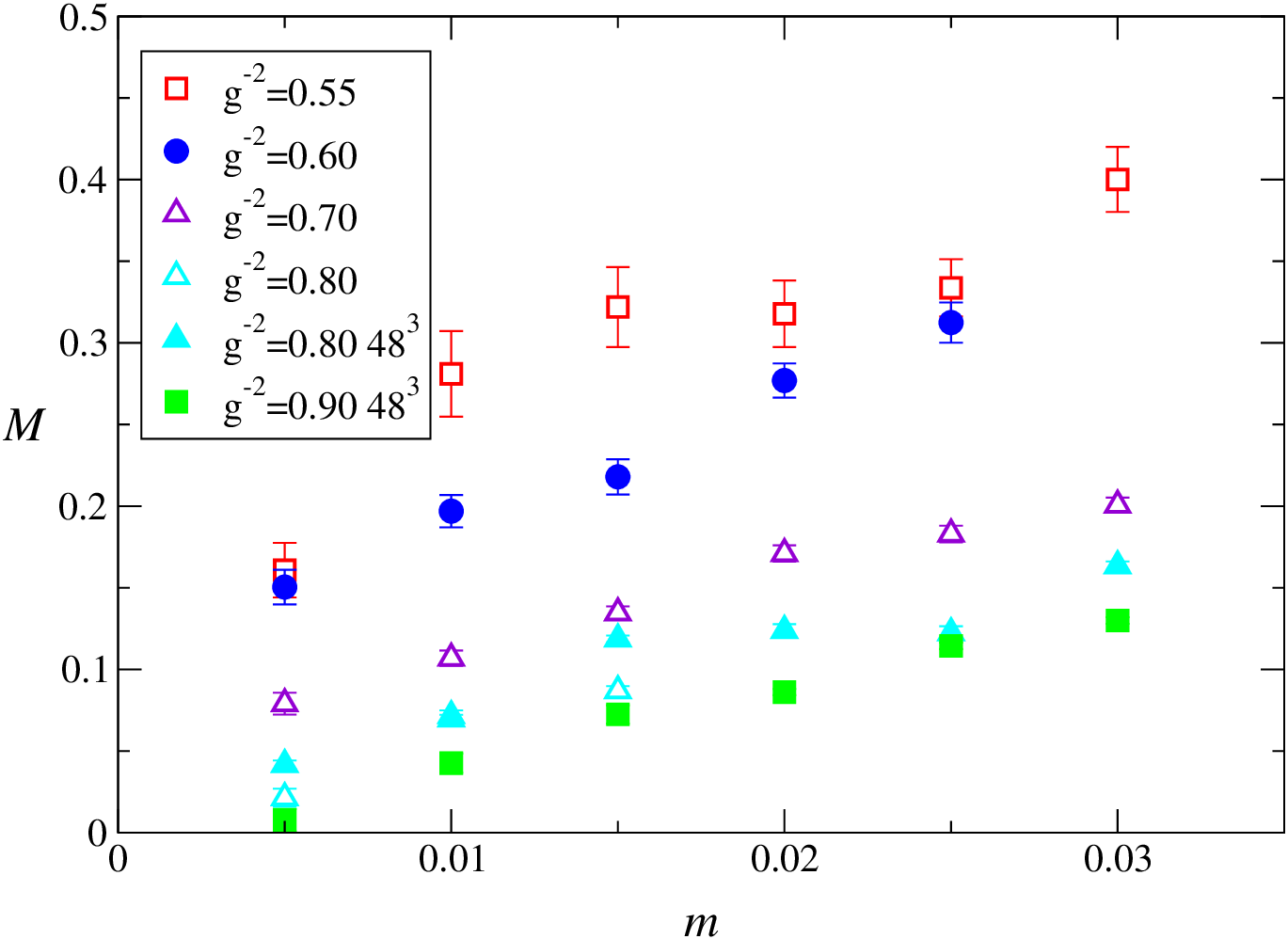}
    \caption{(Color online) 
The fitted parameter $M$ vs. $m$ for various $g^{-2}$.}
   \label{gr:M}
\vspace{8mm}
    \centering
    \includegraphics[width=13.0cm]{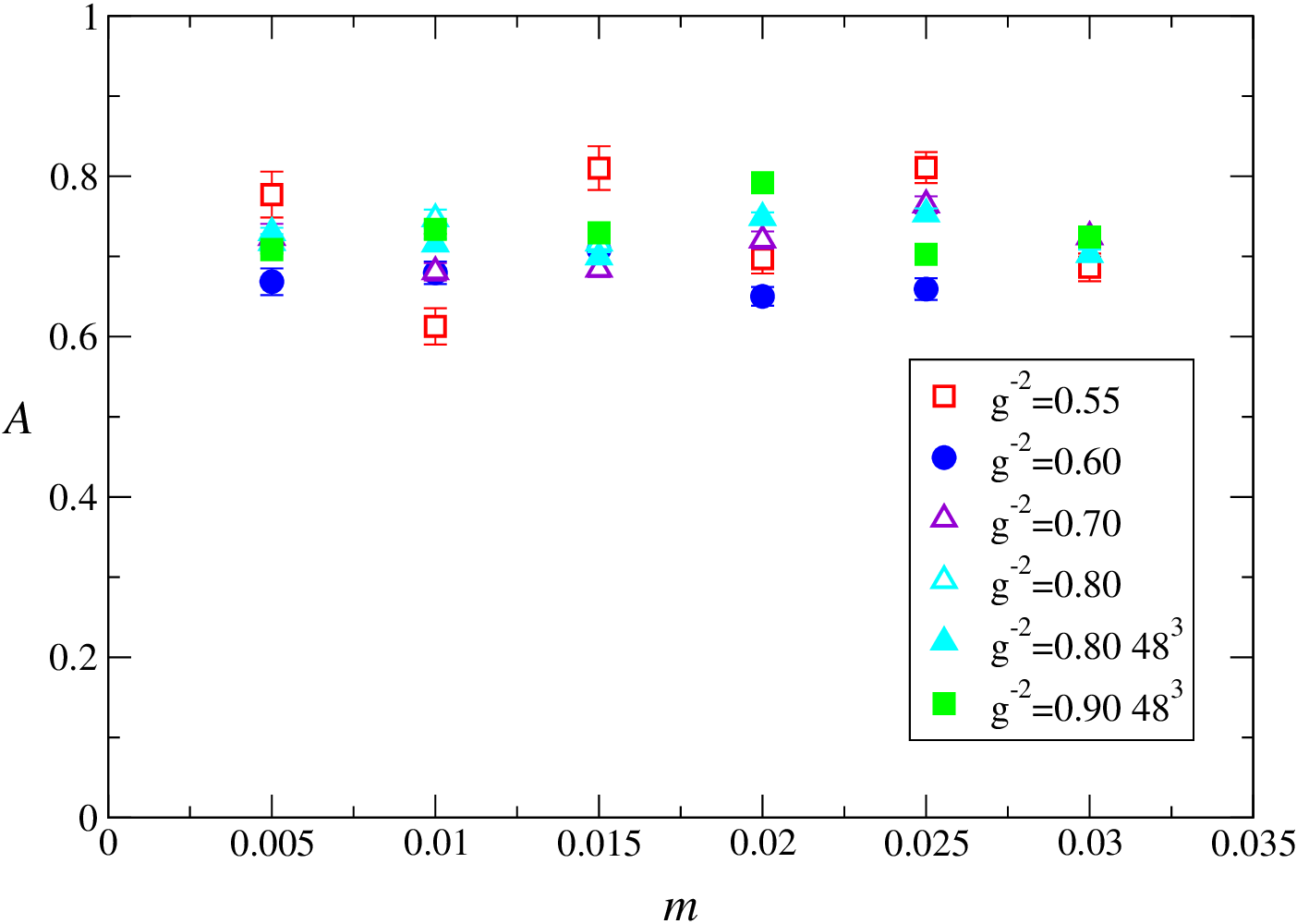}
    \caption{(Color online) 
The fitted parameter $A$ vs. $m$ for various $g^{-2}$.}
   \label{gr:A}
\end{figure*}

The results for $M$ are broadly consistent with our identification of the
critical coupling. For $g^{-2}<g_c^{-2}\simeq0.6$, Fig.~\ref{gr:M} supports
$\lim_{m\to0}M\not=0$, signalling the generation of a gap via spontaneous chiral
symmetry breaking. For
weaker couplings the data can be plausibly extrapolated in the same limit to
$M=0$, signalling a chirally symmetric, conducting phase. Note that throughout the
critical region $\Delta\gg m$, indicating large mass
renormalisation due to strong interactions even in the symmetric phase.
In Sec.~\ref{sec:z} below we will present further results for $M$ for a range of
$m$ in the
critical region.

Fig.~\ref{gr:A} shows that despite some noise in the data the parameter $A$, and
hence the physical Fermi velocity $v_{FR}$, is both $m$- and $g^{-2}$-independent
in the critical region, taking a numerical value $\approx0.7$. We interpret this
as being due to a renormalisation of the bare Fermi velocity $v_{F}\equiv1$ due
to quantum effects. 
This in principle needs be taken into account when we attempt to assign a
physical value to the critical coupling $g_c^{-2}$ in Sec.~\ref{sec:discussion}.
This result is intersting because analytic calculations based on weak-coupling
and/or large-$N_f$ predict that $v_{FR}>v_F$~\cite{vF}.

\subsection{Dynamical Critical Exponent}
\label{sec:z}

In this section we take a closer look at correlations in the critical system,
both via a high statistics study (typically several thousand HMC trajectories)
at $g^{-2}=0.6$, close to the critical value reported in
(\ref{eq:fits}), and a refined study of the quasiparticle mass parameter $M$ in
the critical region. 
All results are from simulations on $24^2\times48$ lattices.
Because the model (\ref{eq:latt}) treats space and time differently,
correlation lengths defined in spatial and temporal directions
can exhibit different critical scaling, leading to two distinct exponents 
defined via~\cite{BW}
\begin{equation}
\xi_s\propto\vert t\vert^{-\nu_s};\;\;\;
\xi_t\propto\vert t\vert^{-\nu_t}.
\label{eq:xi}
\end{equation}
Our goal in this section is to constrain the value of the dynamical
critical exponent $z\equiv\nu_t/\nu_s$
relating spacelike to timelike correlations.

\begin{figure}[htbp]
    \centering
    \includegraphics[width=13.0cm]{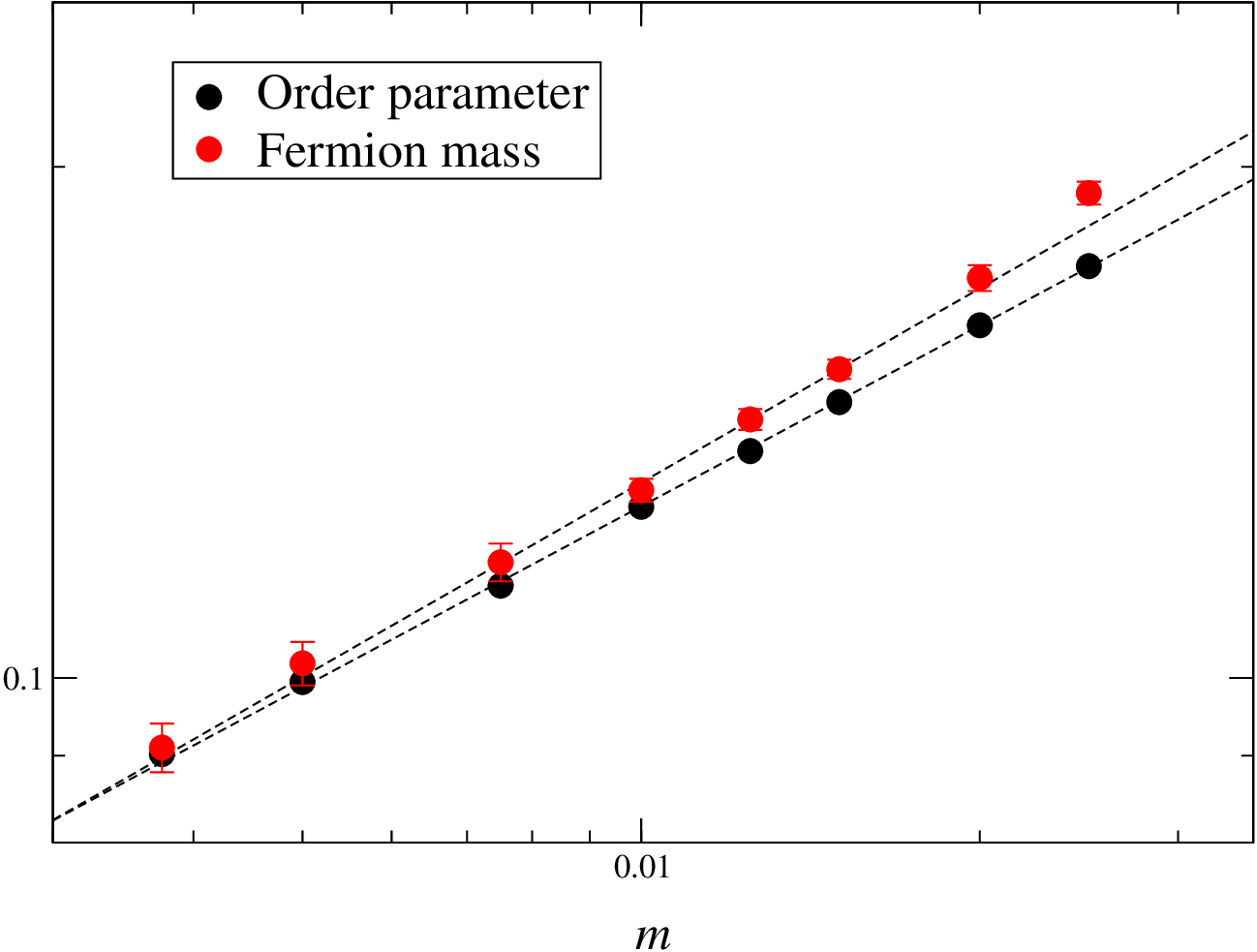}
    \vspace{2mm}
    \caption{(Color online) 
$\langle\bar\chi\chi(m)\rangle$ and $M(m)$ at $g^{-2}=0.6$ on $24^2\times48$.}
   \label{gr:lnln}
\vspace{8mm}
    \centering
    \includegraphics[width=13.0cm]{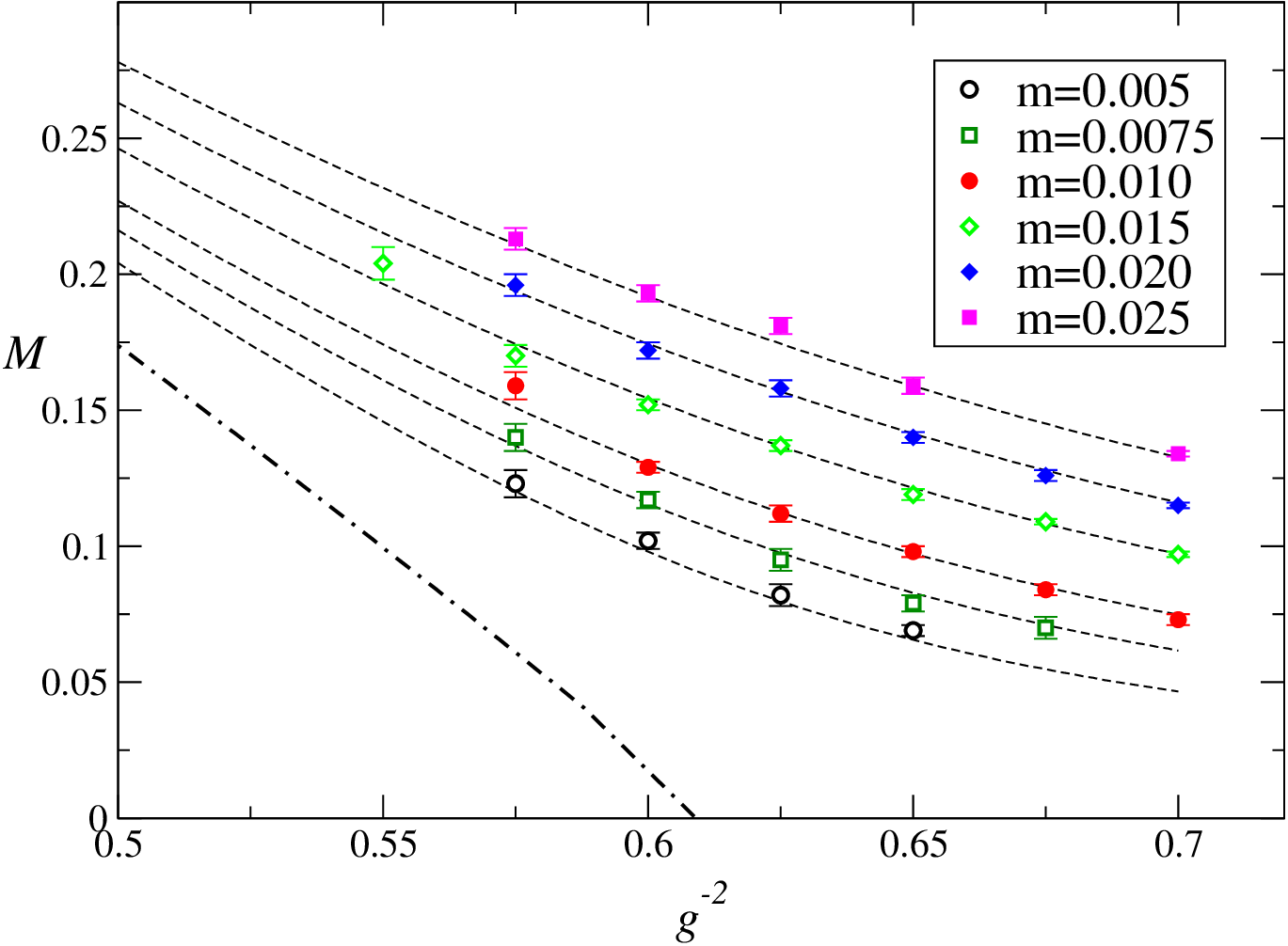}
    \caption{(Color online) 
Fit to (\ref{eq:massfit}) for $M(m)$ data taken on $24^2\times48$.}
   \label{gr:masseosfit}
\end{figure}

Fig.~\ref{gr:lnln} shows data 
for both the order parameter $\langle\bar\chi\chi\rangle$ and mass parameter $M$
defined by (\ref{eq:dispfit}) as a function of $m$ on a log-log plot.
The linear nature of the plots supports a power-law scaling:
\begin{equation}
\langle\bar\chi\chi\rangle\propto m^{1\over\delta};\;\;\; M\propto
m^{\nu_t\over{\delta\beta}},
\label{eq:critscal}
\end{equation}
where $\delta$ and $\beta$ coincide with the definitions implicit in
(\ref{eq:eos}), and the exponent $\nu_t$ is the one relevant 
for the extraction of
spectral properties via (\ref{eq:Efit}) from correlations in the Euclidean 
time direction. 
Least-squares fits (excluding $m=0.025$) 
yield $\delta=2.85(1)$; ${\nu_t\over{\delta\beta}}=0.38(2)$. 
The mismatch between this value for $\delta$ and that of (\ref{eq:fits}) 
extracted from the equation of state is 
ascribed to the actual value of $g_c^{-2}$ lying slightly
above 0.6, as suggested by Fig.~\ref{gr:suscrat}.

Fig.~\ref{gr:masseosfit} shows results for the quasiparticle mass parameter 
$M$ for a range of $m$ and $g^{-2}$ values in the critical region. We have
fitted these data with a relation inspired by the equation of state (\ref{eq:eos}):
\begin{equation}
m=AtM^{{\beta p}\over\nu_t}+BM^{{\delta\beta}\over\nu_t},
\label{eq:massfit}
\end{equation}
which with $M\propto\xi_t^{-1}$ understood recovers (\ref{eq:xi}) in the limit
$m\to0$, and is consistent with (\ref{eq:critscal}) when $t=0$. A fit to 33
datapoints  with $g_c^{-2}$ fixed by (\ref{eq:fits}) yields
\begin{equation}
{{\delta\beta}\over\nu_t}=2.25(5);\;\;\;{{\beta p}\over\nu_t}=1.16(6)
\label{eq:fitnu}
\end{equation}
with $\chi^2$ per degree of freedom of 1.0.

It is now time to discuss the possible anisotropy at
$g^{-2}=g_c^{-2}$ in more detail. As mentioned above, 
the ratio $z=\nu_t/\nu_s$ defines the {\em dynamical critical exponent}, which
is an important characteristic of a QCP. In particular, the
critical dispersion relation is modified to be of the form $E\propto p^z$, which
has important implications for the stability of quasiparticles; energy and
momentum conservation make it  
impossible, in an inelastic collision, for a quasiparticle to decay into 
constituents with smaller $E$ and $p$
if $z<1$~\cite{Son:2007ja}. 

The results in this section permit an estimate of $z$ via the following
indirect argument. First, we use the exponent values from
(\ref{eq:fitnu}) and $\delta,\beta$ from (\ref{eq:fits}) to estimate
\begin{equation}
\nu_t=0.80(3).
\end{equation}
Next, we use a modified hyperscaling relation~\cite{HLS,BW} 
\begin{equation}
\nu_t+(d-1)\nu_s=\beta(\delta+1),
\end{equation}
where $d=3$ is the number of spacetime dimensions, to estimate 
\begin{equation}
\nu_s=0.89(3),
\end{equation}
leading to
\begin{equation}
z=0.90(5).
\end{equation}
This result is tantalising, since although it hints at $z<1$ it 
eliminates neither the value $z\approx0.8$ based on a leading order large-$N_f$
calculation in the strong-coupling limit~\cite{Son:2007ja}, 
nor the general result $z\equiv1$ claimed for systems at a QCP with $d<4$ 
interacting via a
Coulomb potential~\cite{Herbut}. 

\section{Discussion}
\label{sec:discussion}

The main result of this paper is that by numerical means we have identified a quantum critical
point, corresponding to a semimetal-insulator transition,
for a model with $N_f=2$ flavors of Dirac fermion sharing many
symmetries with a low energy effective theory of monolayer graphene. We have
been able to identify critical exponents characterising the transition, as
summarised in eqn.~(\ref{eq:fits}); the most
robust prediction, emerging from a fit to the equation of state, 
and supported by both a  
calculation of the susceptibility ratio $\chi_\ell/\chi_t$ and a direct study of
the scaling of the order parameter against $m$ at $g^{-2}\approx g_c^{-2}$, is that the
exponent $\delta=2.66(3)$. This is significantly different from the value
$\delta=5.5(3)$ obtained in the strong-coupling limit at
$N_f=N_{fc}=4.8(2)$~\cite{Hands:2008id}, demonstrating that the
universality class the model falls into is $N_f$-dependent. A similar picture
has emerged from numerical simulations of the 2+1$d$ Thirring
model~\cite{DelDebbio:1997dv,Christofi:2007ye}, where it has been shown that
$\delta$ {\em increases\/} with $N_f$. Drut and L\"ahde have reported the same
trend from numerical simulations of their model with $N_f=0,2,4$~\cite{D&L3}.
However, their most recent value for $\delta(N_f=2)=2.26(6)$ is significantly
different from ours, so it remains unclear whether the two models lie in the
same universality class, or whther the long range interaction present in the
model of \cite{D&L3} but not here have a decisive effect. 

We have also for the first time presented results for quasiparticle propagation,
finding evidence for a gap developing spontaneously in the spectrum for
$g^{-2}<g_c^{-2}$ as $m\to0$. In addition, analysis of correlations at non-zero
momentum has enabled us to roughly calculate the renormalisation of the
Fermi velocity $v_F$. We reiterate that in our model the fermion propagator is
uniquely defined and readily calculable; in the original model (\ref{eq:model})
the presence of a local gauge symmetry makes analysis of quasiparticle
propagation potentially problematic both
theoretically and numerically. 
 
We have also outlined a method to obtain the
dynamical critical exponent $z$, an important characteristic of any QCP,
using scaling and hyperscaling arguments. Unfortunately the inevitable
accumulation of errors in such an indirect approach precludes us at this stage
from conclusively deciding whether $z<1$ or not. This issue is of theoretical
interest since there are general arguments to claim $z$ is exactly one for
Coulombic systems~\cite{Herbut} (of course, strictly our model is not in this
class). 
In this respect a more direct attempt to
extract $z$ via measurements of the quasiparticle dispersion $E(p)$ on lattices
with a large spatial dimension giving enhanced momentum resolution may prove 
interesting.

Finally, while our model should be regarded as sharing 
universal features of graphene in the neighbourhood of some putative fixed
point, and hence at best able to make predictions of critical exponents
and dimensionless ratios of low-energy observables, it is difficult to resist the
temptation to attempt to convert our result $g_c^{-2}(N_f=2)=0.609(2)$
for the critical coupling into a
physical prediction.

First, we must express our result in terms of the continuum model
(\ref{eq:ThirV}). In order to do this, we remind the reader that the
expression for the propagator (\ref{eq:D_Th}) is derived using a regularisation 
which respects current conservation; unfortunately the lattice regularisation
defined by (\ref{eq:latt}) is not of this type. The solution, as outlined in 
\cite{DelDebbio:1997dv}, is to take the strong-coupling limit of the lattice
model not at $g^{-2}=0$
but at $g^{-2}=g_{\rm lim}^{-2}$, which may be identified numerically via the
location of a peak in $\langle\bar\chi\chi\rangle$~\cite{Christofi:2007ye}.
The relation between the $V$-propagator on the lattice and in the continuum is
then~\cite{DelDebbio:1997dv}
\begin{equation}
D_{\rm latt}(p;g)=ZD_2^\prime(p;g_R)=Z\left[{1\over2g_R^2}+{N_f\over8}{{\vert\vec
p\vert^2}\over(p^2)^{1\over2}}\right]^{-1},
\label{eq:D_latt}
\end{equation}
with $Z=(1-{g^2\over g_{\rm lim}^2})^{-1}>1$ and $g_R^2=Zg^2$.
The extra factor of 2 in the first term in square brackets results from a careful counting
of the staggered fermion degrees of freedom in (\ref{eq:latt}).

For our graphene model Fig.~1 of Ref.~\cite{Hands:2008id} suggests $g_{\rm
lim}^{-2}=0.30(2)$, yielding a renormalised critical coupling
$g_{cR}^2\simeq3.26$.
Now, in order to compare the quantum and classical terms in $D_{\rm latt}$ to
define an effective value of $\lambda$, 
a momentum scale $p$ is needed. Since the only length scale in the problem is the
lattice spacing, a natural (if somewhat arbitrary) choice is
$p_0=0$, $\vert\vec p\vert={\pi\over2}$; this means that the propagators $D_1$
and $D_{\rm latt}$ match at a
distance of roughly one lattice spacing. 
The matching condition 
$\lambda=g^2_R\pi/4$ yields $\lambda_c\approx2.6$. 

Since $D_2$ decays faster than $D_1$ at large distances, this estimate for
$\lambda_c$ is likely to be on the high side. We should however note two factors
neglected in this simplified approach. Firstly, the renormalisation factor
$Z\approx2.0$ boosts the interaction strength of the lattice model; taking
proper account of this will have the effect of raising the predicted $\lambda_c$.
Secondly, the Fermi velocity $v_F$ appearing in (\ref{eq:model}) and
implicitly in (\ref{eq:ThirV}) is the bare one, whereas presumably it is the
renormalised $v_{FR}\approx0.7v_F$ which has the experimental value
$10^6{\rm ms}^{-1}$. Since $\lambda$ in (\ref{eq:lambda}) is defined in terms of
the bare value, this correction has the effect of lowering  the predicted 
$\lambda_c$, although it will also correct the $\lambda$-values calculated for
physically-realised cases such as graphene which is either freely-suspended or
mounted on a substrate of known dielectric constant.
In our view the uncertainty over the phenomenologically-relevant value of $v_F$
must ultimately be settled by an {\it ab initio\/} microscopic calculation;
moreover, should it prove to be the case that $z<1$, then the very notion of a 
universal Fermi velocity becomes ill-defined since $\lim_{p\to0}{dE\over dp}$ diverges.

\begin{table}[h]
\centering
\setlength{\tabcolsep}{0.4pc}
\hspace{-20mm}
\begin{tabular}{|l|lll|}
\hline
reference & $\alpha_c$ & $\lambda_c$ & $N_{fc}$ \\
\hline
this work, \cite{Hands:2008id} & - & 2.6 & 4.8(2) \\
\cite{D&L1,D&L3} & 1.11(6) & 1.70(8) & 4 -- 6 \\
\hline
\cite{Drut:2007zx} & - & - & 2.03\\
\cite{Khveshchenko1,Gorbar:2002iw} & 2.33 & 3.66 & 2.55 \\
\cite{Khveshchenko} & 1.13 & 1.77 & 3.6 \\
\cite{Liu} & 1.16 & 1.82 & 3.5 \\
\cite{Gamayun} & 1.62 & 2.54 & 2.8 \\
\hline
\end{tabular}
\caption{Predictions for the critical interaction strength for $N_f=2$, and for
the critical number of flavors $N_{fc}$ in the strong coupling limit.}
\smallskip
\label{tab:alphac}
\end{table}
Table~\ref{tab:alphac} compares our estimate for the critical interaction
strength with both that of the alternative simulation of Drut and
L\"ahde~\cite{D&L1,D&L2,D&L3}, with a value predicted using a renormalisation
group treatment of radiatively-induced four-fermion contact
interactions~\cite{Drut:2007zx}, 
and with one older and three more recent
estimates~\cite{Khveshchenko1,Gorbar:2002iw,Khveshchenko,Liu,Gamayun} based on
self-consistent diagrammatic calculations using the model (\ref{eq:model}).
Note that is conventional in the literature to quote the critical effective fine
structure constant $\alpha_c=4\lambda_c/\pi N_f$; here we show both parameters
where appropriate. Also listed are
estimates for the critical number of flavors $N_{fc}$ corresponding to the
location of the QCP in the strong-coupling limit. We argued in \cite{Hands:2008id}
that in this limit our model coincides with (\ref{eq:model}) and hence that
$N_{fc}=4.8(2)$ is a robust non-perturbative prediction. 

To give these numbers
some meaning recall that $\lambda$ is calculated to be 1.25 for graphene 
on a SiO$_2$ substrate, where experimentally it is known to be a conductor, and
3.4 in vacuum. Acknowledging the difficulties in obtaining precise numbers
reviewed in the previous paragraphs, we may nonetheless observe that our
simulations lend support to the mounting body of theoretical evidence that
freely-suspended graphene is an insulator.

\section{Acknowledgements}

This project required approximately 1,000,000 core hours to complete.
The cpus used were either Intel(R) Xeon(R) E5420 or Dual-Core AMD
Opteron(tm) 2218 in a dual configuration (2x2 cores).
The authors wish to thank Diamond Light Source for kindly allowing them
to use extensive computing resources, and specifically Tina Friedrich,
Frederik Ferner and Alun Ashton for help in configuring and maintaining
these resources.
WA would like to thank Gwyndaf Evans for support and kind words of
encouragement.

\end{document}